\documentclass[twocolumn,showpacs,preprintnumbers,superscriptaddress,prl,floatfix,aps,10pt]{revtex4-2}
\usepackage[utf8]{inputenc}
\usepackage{natbib}
\usepackage{amsmath,amssymb}
\usepackage{xcolor}
\usepackage{graphicx}
\usepackage{url}
\usepackage{hyperref}
\hypersetup{
    colorlinks=true,
    citecolor=blue,
    linkcolor=blue,
    filecolor=magenta,      
    urlcolor=blue,
    pdfpagemode=FullScreen,
    }
\renewcommand {\vec}    [1]    {\ensuremath{\boldsymbol{#1}}}
\newcommand   {\avg}    [1]    {\ensuremath{\left\langle#1\right\rangle}}

\makeatletter

\begin{document}

\title{Microscopic theory of phonon polaritons and long wavelength dielectric response}
\date{\today}

\author{Olle Hellman}
\affiliation{Department of Molecular Chemistry and Materials Science, Weizmann Institute of Science, Rehovoth 76100, Israel}

\author{Leeor Kronik}
\affiliation{Department of Molecular Chemistry and Materials Science, Weizmann Institute of Science, Rehovoth 76100, Israel}

\begin{abstract}
We present a first-principles approach for calculating phonon-polariton dispersion relations. In this approach, phonon-photon interaction is described by quantization of a Hamiltonian that describes harmonic lattice vibrations coupled with the electromagnetic field inside the material. All Hamiltonian parameters are obtained from first-principles calculations, with diagonalization leading to non-interacting polariton quasiparticles. This method naturally includes retardation effects and resolves non-analytical behavior and ambiguities in phonon frequencies at the Brillouin zone center, especially in non-cubic and optically anisotropic materials. It also allows for the natural emergence of hyperbolic polaritons. Furthermore, by incorporating higher-order terms in the Hamiltonian, we also account for quasiparticle interactions and spectral broadening. Specifically, we show how anharmonic effects in phonon polaritons lead to a dielectric response that challenges traditional models. The accuracy and consequences of the approach are demonstrated on GaP and GaN as harmonic test systems and PbTe and $\beta$-Ga$_2$O$_3$ as anharmonic test systems. 
\end{abstract}

\maketitle

Phonon polaritons -- hybrid quasiparticles resulting from strong coupling between optical phonons and electromagnetic fields -- have been a subject of both fundamental and applied interest since their theoretical description by \citet{Huang.1951,Huang.1951b} and subsequent experimental confirmation by \citet{Henry.1965} in the mid-20$^{\rm th}$ century.
In the decades following the seminal discovery,  theoretical description of phonon polaritons was primarily confined to phenomenological models applicable to diatomic and isotropic crystals \cite{bradnmuller1975,Giallorenzi.1971,Benson.1970,Pick.1970}.
However, rapid progress in experimental techniques has recently revitalized theoretical interest in these mixed excitations \cite{Sellati.2025}. Modern investigations have revealed rich phenomena, including nontrivial dispersion relations in anisotropic and complex crystal structures and pronounced anharmonic effects \cite{Passler.2022,Pan.2023,Matson.2023,Bergeron.2023}. These phenomena  demand explanation in terms of a rigorous first-principles treatment of light-matter coupling . At optical frequencies of light the theory for light-matter interactions is well developed \cite{Ruggenthaler.2023}, but here we focus on coupling to the ions, which drives the phonon-polariton picture, rather than on coupling between electrons and the electromagnetic field.

Here, we present a unified Hamiltonian framework in which both lattice vibrations and the electromagnetic field are treated as dynamical degrees of freedom. Starting from a generalized Coulomb-gauge formulation, we derive a quadratic Hamiltonian that naturally incorporates the coupling between phonons and photons, leading to the formation of phonon polaritons. Our approach not only recovers well-established results in the harmonic limit, thereby providing a nonempirical validation of traditional theories, but also explains new observations associated with anisotropy and facilitates the systematic inclusion of anharmonic interactions. This is particularly important for understanding phenomena such as hyperbolic polaritons \cite{Dai.2014,Galiffi.2024} that arise in realistic materials at finite temperatures, as well as linewidth broadening, deviations from ideal black-body radiation, and complex dielectric responses. 

Our starting point is the construction of a Hamiltonian for harmonic particles interacting with an electromagnetic (EM) field that obeys Maxwell's equations, in a generalized Coulomb gauge, $\nabla \cdot (\vec{\epsilon} \vec{A}) = 0$, where $\vec{\epsilon}$ is the high-frequency dielectric tensor and $\vec{A}$ is the EM vector potential (Hartree atomic units are used throughout):

\begin{equation}
\begin{split}
    \label{eq:startinghamiltonian}
	H = 
    &
    \sum_i \frac{ m_i \dot{\vec{u}}_i^2
    }{
        2 
    } 
    +     
    \frac{1}{2} 
    \sum_{ij}
    \vec{u}_i^T \vec{\Phi}_{ij} \vec{u}_j
    +
\\  
    &
    +
    \frac{1}{2} 
    \sum_{ij}
    \vec{u}_i^T \vec{Z}_i^T
    \vec{\epsilon}^{-1}
    \vec{Z}_j \vec{u}_j
\\    
    & +
    \int 
    \left[
	\frac{1}{2} (\nabla \times \vec{A}) \vec{\mu} (\nabla \times \vec{A}) + 
    \right.
 \\
    & +
    \left.
    \frac{1}{2} \vec{\dot{A}} \vec{\epsilon} \vec{\dot{A}}
    - \vec{A} \cdot 
        \sum_i \vec{Z}_i \dot{\vec{u}}_i \delta(\vec{r}-\vec{u}_i)
    \right]    
    dr    
    .
\end{split}
\end{equation}
The first two terms in the above equation describe the standard harmonic phonon Hamiltonian, where $\vec{u}_i$ is the displacement of atom $i$, $m_i$ the corresponding mass, and $\vec{\Phi}_{ij}$ are force constants between atoms $i$ and j. The next term describes screened Coulomb dipole interactions by considering the Born effective charges, $\vec{Z}_i$, along with $\vec{\epsilon}$. The next two terms correspond to magnetic and electric ($B^2$ and $E^2$) energies in the electromagnetic Hamiltonian, expressed in terms of $\vec{A}$. The final term describes dipole interactions with the EM field, $\vec{A}\cdot\vec{j}$, where $\vec{j}$ is the total (bound) current density. Note that $\vec{Z}_i\dot{\vec{u}_i} = \dot{\vec{P}_i} = \vec{j}_i$, where $\vec{P}_i$ is the polarization originating from atom $i$. For a detailed derivation of the equations of motion corresponding to the Hamiltonian of Eq.\ \eqref{eq:startinghamiltonian},  see Section I of the Supplementary Information (SI).

To quantize the system, we introduce the non-interacting phonon Hamiltonian~\cite{born1954dynamical}
\begin{equation}
    H^{\textrm{phonon}} = 
    \sum_{\vec{q}s} \frac{\omega_{\vec{q}s}}{4}
    \left(
        A^{\dagger}_{\vec{q}s} 
        A_{\vec{q}s}
        +
        B^{\dagger}_{\vec{q}s}
        B_{\vec{q}s} 
    \right)\, ,
\end{equation}
where $\omega_{\vec{q}s}$ is the phonon frequency at wavevector $\vec{q}$ and mode $s$. The operators $A$ and $B$ are linear combinations of the creation and annihilation operators for a harmonic oscillator and represent dimensionless position and momentum, respectively -- see Section II of the SI for details. We also introduce the non-interacting photon Hamiltonian~\cite{zangwill2013modern}:
\begin{equation}
    \label{eq:photonhamiltonian}
    H^{\textrm{EM}} =
    \sum_{\vec{q}r} \frac{\omega_{\vec{q}r}}{4}
    \left(
        \mathcal{E}^{\dagger}_{\vec{q}r} 
        \mathcal{E}_{\vec{q}r}
        +
        \mathcal{A}^{\dagger}_{\vec{q}r}
        \mathcal{A}_{\vec{q}r} 
    \right)\,,
\end{equation}
where $\omega_{\vec{q}r}$ is the photon frequency at wavevector $\vec{q}$ and transverse branch $r$.  $\mathcal{E}$ and $\mathcal{A}$ are linear combinations of the photon creation and annihilation operators, representing dimensionless electric field and vector potential, respectively -- see Section III of the SI for details. Note that as a consequence of the Coulomb gauge choice we do not include the longitudinal part of the electromagnetic field in Eq.\ \eqref{eq:photonhamiltonian}, only the two transverse branches are present.

We can now express Eq.\ \eqref{eq:startinghamiltonian} in terms of the non-interacting phonon and photon operators. Complete algebraic details are given in Section IV of the SI. The result for the coupling terms in the Hamiltonian (third and sixth term in Eq.\ \eqref{eq:startinghamiltonian}) is:
\begin{equation}    
    H^{\textrm{mix}} = \sum_{\vec{q}ss'} X_{\vec{q}ss'} A^\dagger_{\vec{q}s}A_{\vec{q}s}
    +
    \sum_{\vec{q}rs} Z_{\vec{q}rs} A^\dagger_{\vec{q}s} \mathcal{E}_{\vec{q}r}
\end{equation}
where
\begin{align}
    X_{qss'} & =
    \sum_{ij\alpha\beta\mu\nu}
     \frac{
        \epsilon_{\vec{q}s}^{i\alpha}
    Z^{\mu}_{i\alpha}
    (\epsilon^{-1})_{\mu\nu}
    Z^{\nu}_{j\beta}
    \epsilon_{\vec{q}s'}^{j\beta}
        }{
     \sqrt{
        m_i m_j \omega_{\vec{q}s} \omega_{\vec{q}s'}
        }
    } 
\\    
    Z_{qrs} & =
    \sum_{i\alpha\mu}
    \frac{1}{2}
    \sqrt{ \frac{\omega_{\vec{q}r}}{m_i \omega_{\vec{q}s}}}
    v_{\vec{q}r}^{\mu}
    Z^{\mu}_{i\alpha}
    \epsilon_{\vec{q}s}^{i\alpha}.        
\end{align}
Here $\epsilon_{\vec{q}s}$ and $v_{\vec{q}r}$ are the phonon and the photon eigenvectors and $\alpha, \beta, \mu, \nu$ are cartesian directions.  Physically, $X$ represents electromagnetic interactions arising from the electric field created as a phonon is excited, $Z$ describes coupling between ionic motion and the electric field.

The total Hamiltonian can be expressed as $H^{\textrm{lattice}} + H^{\textrm{EM}} + H^{\textrm{mix}}$ and is quadratic in operators. Therefore, we seek a linear combination (transformation matrices $\vec{U}$ and $\vec{V}$  of phonon and photon operators, respectively) that diagonalizes the coupled Hamiltonian, leading to  harmonic quasiparticles that are linear combinations of the phonons and photons, i.e., phonon polaritons. Details are given in Section V of the SI. Briefly, we define a transformation 
\begin{align}
    K_{\vec{q}j} = 
    \sum_{s} \vec{U}_{\vec{q}js} A_{\vec{q}s} + 
    \sum_{r} \vec{U}_{\vec{q}jr} \mathcal{E}_{\vec{q}r},
\\
    L_{\vec{q}j} = 
    \sum_{s} \vec{V}_{\vec{q}js} B_{\vec{q}s} + 
    \sum_{r} \vec{V}_{\vec{q}jr} \mathcal{A}_{\vec{q}r},
\end{align}
where $j$ is a polariton branch. We choose $\vec{U}$ and $\vec{V}$ so as to diagonalize the total Hamiltonian (\ref{eq:startinghamiltonian}) and find them in practice by requiring that the new quasiparticles obey the harmonic commutation relations~\cite{Sakurai_Napolitano_2020}:
\begin{equation}
\label{eq:canonicalcommutation}
\begin{split}
    \vec{\xi}^2 \vec{K}_{\vec{q}} & = [[\vec{K}_{\vec{q}},\vec{H}],\vec{H}],
\\
    \vec{\xi}^2 \vec{L}_{\vec{q}} & = [[\vec{L}_{\vec{q}},\vec{H}],\vec{H}],
\end{split}
\end{equation}
where $\vec{\xi}$ is a diagonal matrix with polariton frequencies on the diagonal. Utilizing the commutation relations of the pure phonon and photon operators allows us to express Eq. \eqref{eq:canonicalcommutation} as
\begin{align}
    \begin{pmatrix}
        \vec{\omega}^p & 0 \\
        0 & \vec{\omega}^{em}
    \end{pmatrix}
    \begin{pmatrix}
        \vec{\omega}^p + 4\vec{X} & 2\vec{Z} \\
        2\vec{Z}^\dagger & \vec{\omega}^{em}        
    \end{pmatrix}
    \vec{U}
    & =
    \vec{\xi}\vec{U},
\end{align}
i.e., as an eigenvalue equation that determines $\vec{U}$, with $\vec{V}=\vec{U}^{-1}$. Solving this eigenvalue problem leads to a phonon-polariton basis with a quadratic Hamiltonian given by:
\begin{equation}
    \label{eq:barepolaritonH}
    H = \sum_{\vec{qj}}
    \frac{\xi_{\vec{q}j}}{4}
    \left(
        K^\dagger_{\vec{q}j}K_{\vec{q}j} + 
        L^\dagger_{\vec{q}j}L_{\vec{q}j}
    \right) \,
\end{equation}
At every wavevector $\vec{q}$, we have $3N + 2$ polariton frequencies $\xi$ (where $N$ is the number of atoms per unit cell) -- $3N$ from the phonons and two from the photons. Applying Eq.\ \eqref{eq:barepolaritonH} to a specific material, however, requires knowledge of the interactions defined in Eq. \eqref{eq:startinghamiltonian}. The parameters that are needed are the interatomic force constants $\vec{\Phi}_{ij}$, the Born effective charges $\vec{Z}_i$, and the high-frequency dielectric tensor $\vec{\epsilon}$. All of these are readily available, without empiricism, from perturbation theory based on first-principles density functional theory (DFT) calculations \cite{Baroni.2001} -- see Section VI of the SI for computational details. Here we use the temperature-dependent effective potential (TDEP) approach \cite{Hellman.2013,Hellman.2013p9g,Hellman.2011,Knoop.2024} to determine all interaction parameters in a manner that includes renormalization to finite temperature.

\begin{figure}[ht!]
    \centering
    \includegraphics{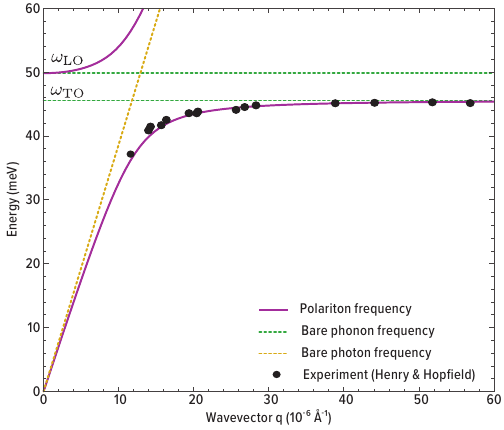}    
    \caption{\label{fig:gap}Phonon-polariton dispersion relation in GaP. Polariton frequencies obtained from eigenvalues of Eq.\ (\ref{eq:barepolaritonH}) are compared with experimental values from \citet{Henry.1965}, as well as with bare phonon and photon frequencies.}
\end{figure}

\begin{figure*}[ht!]
    \centering
    \includegraphics[width=\linewidth]{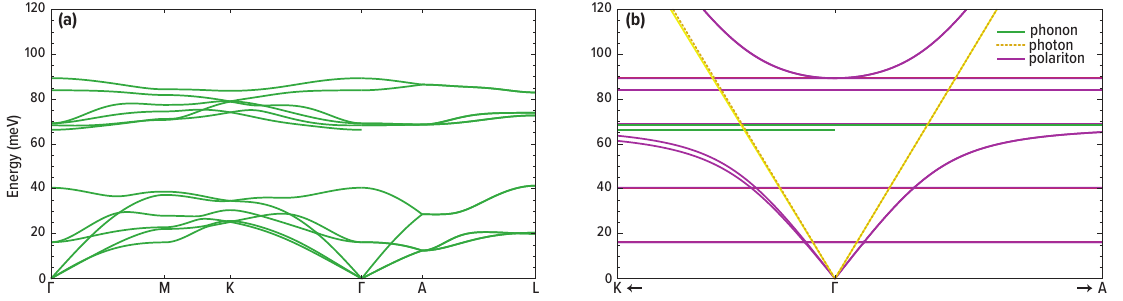}
    \caption{\label{fig:gan_polariton}
    (a) Phonon dispersion relations for wurtzite GaN. Note the discontinuity at the zone center, $\Gamma$, for the optical modes. (b) Dispersion relations for bare photons (yellow), bare phonons (green), and phonon-polaritons (purple) in a zoomed-in region close to the zone center. No discontinuities arise in the polariton picture and the Hamiltonian is analytical throughout the Brillouin zone.}
\end{figure*}

To verify the utility and accuracy of the above considerations, we apply it to GaP, for which phonon-polaritons were first measured ~\cite{Henry.1965}. The results, given in Fig.\ \ref{fig:gap}, demonstrate excellent agreement with experimental results. They also agree perfectly with older theory, where dispersion relations were fitted based on the Lyddane-Sachs-Teller relations~\cite{Lyddane.1941,Huang.1951b}. However, the results  presented here are obtained entirely from first principles and are in no way limited to a specific space group or structure. In particular, it therefore allows for the prediction of hyperbolic polaritons, an issue we return to below after we introduce appropriate broadening.

As a first demonstration of the advantages of this more general treatment, we apply the method to wurtzite GaN, which is optically anisotropic, such that the traditional way of determining polariton dispersion relations breaks down. The results are given in Fig.\ \ref{fig:gan_polariton} and allow for interesting observations relating to the long-range behavior of phonon frequencies. In non-cubic systems, there are generally non-analytical components to the dynamical matrix at the zone center at the bare phonon level~\cite{born1954dynamical}. This leads to some conceptual difficulties. Consider the phonon ground state energy. This quantity is ill-defined, because at the zone center it depends on the direction in which the probe is applied. And yet the ground state energy should not have any implicit probe associated with it. As shown in Fig.\ \ref{fig:gan_polariton}, all ambiguities are removed in the polariton picture.  

The above-discussed removal of singularities can intuitively be explained as a consequence of retardation. In a non-relativistic phonon picture, an optical mode corresponds to displacement of all atoms in each unit cell at the same time. This leads to the formation of a macroscopic dipole in the system, causing lower vibrational frequencies transverse to the field (TO phonons) and higher frequencies along it (LO phonons).
This simple picture changes if we include retardation effects. In a given unit cell, the electric field due to the induced dipole from another unit cell possesses a time lag, such that each neighboring cell appears slightly out of phase. This hinders the buildup of a macroscopic dipole, and there is no macroscopic field to be transverse or longitudinal to. Hence, no splitting of the optical phonons at the zone center is expected. Away from the macroscopic limit, i.e., at finite $\vec{q}$ such that wavelength is not long enough for retardation effects, the usual splitting between longitudinal and transverse phonons emerges.

Other than the phonon frequencies, the ionic contribution to the dielectric response, $\vec{\epsilon}_{\rm ion}(\vec{q},\Omega)$, is also directly affected by the polariton picture. The ionic contribution is determined as the Fourier transform of the polarization autocorrelation function~\cite{Cowley.19637cj}:
\begin{equation}
    \label{eq:polpolcorrelation}
    \epsilon^{\mu\nu}_{\rm ion}(\vec{q},\Omega) = \int \avg{P_{\vec{q}}^\mu(t),P_{\vec{q}}^\nu(0)} e^{i\Omega t} dt,
\end{equation}
where $P_{\vec{q}}^\mu$ is the overall polarization associated with wavevector $q$ along the direction $\mu$. In the harmonic phonon picture, this reduces to~\cite{Gonze.1997}:
\begin{equation}    
    \epsilon^{\mu\nu}_{\rm ion}(\vec{q},\Omega)
    =
    \lim_{\eta \rightarrow 0}
    \frac{1}{4\pi}
    \sum_s 
    \frac{Z^\mu_{qs} Z^\nu_{qs}}
    {\omega_{\vec{q}s}^2 - (\Omega+i\eta)^2 },
\end{equation}
where
\begin{equation}Z^\mu_{\vec{q}s}  =
    \sum_{i\alpha}
    \frac{1}{ \sqrt{m_i} }
    \epsilon_{\vec{q}s}^{i\alpha}
    Z_{i\alpha}^\mu.
\end{equation}
In the long-wavelength static limit, for a cubic system with two atoms in the unit cell, this becomes a strictly real quantity, given by
\begin{equation}
\label{eq:phonon_static_eps}
    \epsilon^{\mu\nu}_i(\vec{q}=0,\Omega=0)
    =    
    \frac{1}{4\pi}
    \frac{Z^\mu_{TO} Z^\nu_{TO}}
    {\omega_{TO}^2 } \,.
\end{equation}
With some straightforward algebra, the above equation can be shown to be equivalent to  the Lyddane-Sachs-Teller relation~\cite{Lyddane.1941}.

If we consider the polariton picture, we arrive at a similar expression, but it is subtly different owing to the different effective charge (denoted by tilde) and vibration frequencies ($\xi$)  of the quasi-particles:
\begin{equation}   
    \tilde{\epsilon}^{\mu\nu}_{\rm ion}(\vec{q},\Omega)
    =
    \lim_{\eta \rightarrow 0}
    \frac{1}{4\pi}
    \sum_s 
    \frac{\tilde{Z}^\mu_{qs} \tilde{Z}^\nu_{qs}}
    {\xi_{\vec{q}s}^2 - (\Omega+i\eta)^2 },
\end{equation}
where
\begin{equation}
    \tilde{Z}^\mu_{\vec{q}s} =
    \sum_{i\alpha}
    \frac{1}{ \sqrt{m_i} }
    \nu_{\vec{q}s}^{i\alpha}
    Z_{i\alpha}^\mu \,.
\end{equation}
If we consider this in the static limit for a cubic system with two atoms in the unit cell, we obtain:
\begin{equation}
\label{eq:polariton_static_eps}
    \tilde{\epsilon}^{\mu\nu}_{\rm ion}(\vec{q},\Omega=0)
    =
    \frac{1}{4\pi}
    \sum_s 
    \frac{\tilde{Z}^\mu_{qs} \tilde{Z}^\nu_{qs}}
    {\xi_{\vec{q}s}^2 }
\end{equation}
The difference of the polariton-based Eq.\ \eqref{eq:polariton_static_eps} from the phonon-based Eq.\ \eqref{eq:phonon_static_eps} is that the polariton frequencies $\xi$ and eigenvectors $\nu$ that determine the oscillator strength are still a function of $\vec{q}$ (as shown in Fig.\ \ref{fig:gap}), especially close to the macroscopic limit, and therefore so is $\tilde{\epsilon}_{\rm ion}$. 

\begin{figure}[htb]
    \includegraphics[width=\linewidth]{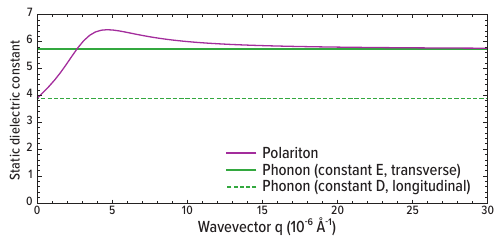}
    \caption{Static dielectric constant for NaCl as a function of wavevector $\vec{q}$, comparing polariton, closed-circuit (constant-E, transverse) and open-circuit (constant-D, longitudinal) limits.
    \label{fig:eps0}
    }
\end{figure}

\begin{figure*}[htb]
    \includegraphics[width=\linewidth]{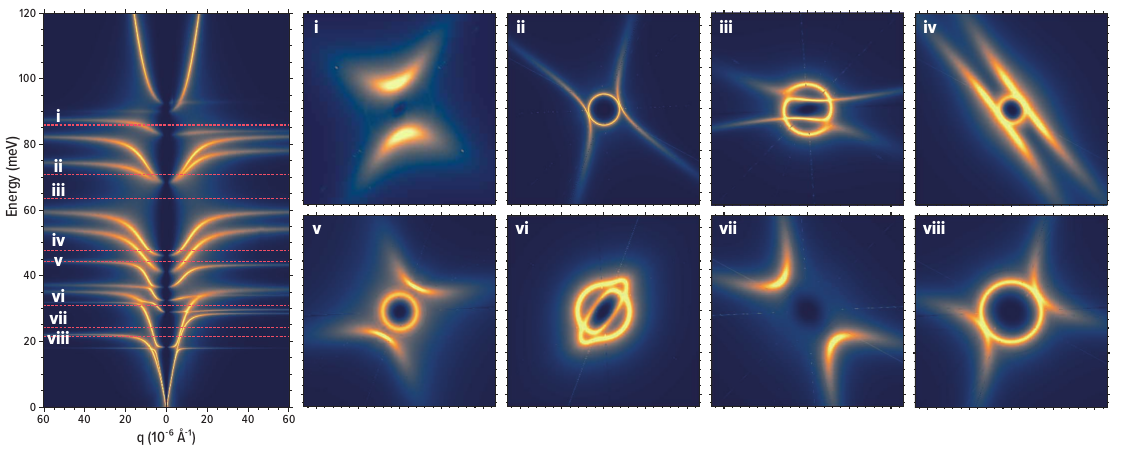}
    \caption{Left panel: Electromagnetic part of the phonon polariton spectral function in $\beta$-Ga$_2$O$_3$. Panels (i)-(viii) show constant energy cuts of the spectral function, where the energies for the respective slice is marked in the left panel. The $\vec{q}$-vectors in each slice lie in the plane orthogonal to $(010)$. 
    \label{fig:hyperbolic}   
    }
\end{figure*}

Fig. \ref{fig:eps0} compares the  phonon and polariton picture for the overall static dielectric constant in NaCl in the small $q$ region. Key differences are observed. In the polariton picture, $\epsilon$ exhibits clear dispersion close to the zone center, and in the true long wavelength limit it is determined by a (triply degenerate in cubic materials) polariton of phonon character at the zone center, with frequency $\xi = \omega_{LO}$. Within the traditional phonon picture, however, the dependence on $q$ in this regime is negligible and the limit is determined by $\omega_{TO}$ (solid green line in the figure).

The duality of the long wavelength dielectric response has previously been explained by \citet{Stengel.2009} as a consequence of different electromagnetic boundary conditions. Under closed-circuit boundary conditions (constant $\vec{E}$-field), longitudinal fields are suppressed and only the transverse response remains, whereas in open-circuit boundary conditions (constant $\vec{D}$-field), the transverse field is suppressed. In Fig. \ref{fig:eps0} we see that the polariton picture smoothly joins the two extremes.
The Lyddane-Sachs-Teller relation recovers the closed-circuit (constant-E) limit, whereas a static capacitor measurement (corresponding to $q=0$) yields the open-circuit (constant-D) limit. One could think of the different limits in terms of screening of long-range interactions. In the polariton picture it is relativistic effects that screen the long-range interactions, whereas in the boundary condition picture the leads are responsible for generating long-range screening. 

Before we return to the issue of hyperbolic polaritons, we note that to facilitate a more direct comparison to experiment, broadening effects should also be taken into account. Historically, broadening associated with finite quasiparticle lifetime has been treated phenomenologically, by assuming a complex dielectric function~\cite{Wright.1969,Benson.1970}. This naturally raises the question of how to introduce broadening parameters to our first-principles theory {\it without} invoking empirical broadening parameters. In our treatment, this can be achieved simply by avoiding truncation of the expansion of the energy in the Hamiltonian of Eq.\ \eqref{eq:startinghamiltonian} at the second order \cite{Roman.2006}. Specifically, the {\em third} order term in the expansion, $H_1$, representing the lowest-order anharmonic effects in the atom displacements and the electric field, is given by: 
\begin{equation}
\begin{split}
\label{eq:anharmonicterms}
    H_1 & = 
    \frac{1}{3!} 
    \sum_{ijk\alpha\beta\gamma}\frac{
        \partial^3 U
    }{
        \partial u_{i\alpha} \partial u_{j\beta} \partial u_{k\gamma}
    } u_{i\alpha} u_{j\beta} u_{k\gamma} + 
\\
    & +
    \frac{1}{3!}
    \sum_{ij\alpha\beta\mu}
    \frac{
        \partial^3 U
    }{
        \partial u_{i\alpha} \partial u_{j\beta} \partial E_{\mu}
    } u_{i\alpha} u_{j\beta} E_{\mu} + 
\\
    & +
    \frac{1}{3!} 
    \sum_{i\alpha\mu\nu}
    \frac{
        \partial^3 U
    }{
        \partial u_{i\alpha} \partial E_{\mu} \partial E_{\nu}
    } u_{i\alpha} E_{\mu} E_{\nu} + 
\\
    & +
    \frac{1}{3!}
    \sum_{i\mu\nu\xi}
    \frac{
        \partial^3 U
    }{
        \partial E_{\mu} \partial E_{\nu} \partial E_{\xi}
    } E_{\mu} E_{\nu} E_{\xi} \,.
\end{split}
\end{equation}
The first term in the above equation is the usual three-body anharmonic term, the second represents the next-order Born charges, the third is a real-space representation of the Raman tensor, and the fourth corresponds to hyperpolarizability. As in the harmonic case, the numerical values of the third-order derivatives that constitute the anharmonic interaction parameters are determined via first principles calculations, with the numerical procedure used to determine them described in detail in Ref.\ \cite{Benshalom.2022}. 

If we now express the displacements $u$ and electric field $E$ in terms of the non-interacting polariton operators of Eq.\ \eqref{eq:barepolaritonH}, we obtain (see Section VII of the SI for details):
\begin{equation}
\begin{split}
    \label{eq:anharmonicpolaritonH}
    H = & \sum_{\vec{qj}}
    \frac{\xi_{\vec{q}j}}{4}
    \left(
        K^\dagger_{\vec{q}j}K_{\vec{q}j} + 
        L^\dagger_{\vec{q}j}L_{\vec{q}j}
    \right) +
\\    
    + &
    \sum_{qq'q''jj'j''}
    \Theta_{qq'q''}^{jj'j''}
    K_{\vec{q}j}K_{\vec{q}'j'}K_{\vec{q}''j''},
\end{split}
\end{equation}
where the first line in the above is the harmonic Hamiltonian of Eq.\ \eqref{eq:barepolaritonH}, and the second line comprises anharmonic three-body interactions, with coefficients  $\Theta$ (with higher-order $L$ terms avoided owing to the neglect of high-order magnetic fields). 

The form of Hamiltonian in Eq.\ \eqref{eq:anharmonicpolaritonH} is amenable to analysis in terms of standard many-body techniques \cite{Leibfried.1961h9,Maradudin.1962,Cowley.19637cj}. Specifically, the polariton spectral function, $J(\Omega$), per wave vector $q$ and modes $j$, $j'$, is given by
\begin{equation}
    \label{eq:spectralfunction}
    J_{\vec{q}jj'}(\Omega) = 
    \lim_{\eta \rightarrow 0} -\frac{1}{\pi} 
    \Im \left\{  G_{\vec{q}jj'}(\Omega + i\eta) \right\},
\end{equation}
where 
\begin{equation}
    G_{\vec{q}jj'}(\Omega)^{-1} = 
    G^0_{\vec{q}jj'}(\Omega)^{-1} + 
    \Sigma_{\vec{q}jj'}(\Omega),
\end{equation}
\begin{equation}
    \label{eq:selfenergy}
    \Sigma_{\vec{q}jj'}(\Omega) = -18 \sum_{\vec{q}_1\vec{q}_2 s_1 s_2}
    \Theta^{j s_1s_2}_{\vec{q}\bar{\vec{q}}_1\bar{\vec{q}}_2}
    \Theta^{j' s_1s_2}_{\bar{\vec{q}} \vec{q}_1 \vec{q}_2}
    S(s_1,s_2,\Omega),    
\end{equation}
and
\begin{equation}
\begin{split}
& S(s_a,s_b,\Omega) = 
\\
&	(n_{a}+n_{b}+1)
	\left[
	   \frac{1}{ (\omega_{a}+\omega_{b}-\Omega)_p }-
	   \frac{1}{ (\omega_{a}+\omega_{b}+\Omega)_p }
	\right]+
	\\
&	 +
	(n_{a}-n_{b})
	\left[
	\frac{1}{ (\omega_{b}-\omega_{a}+\Omega)_p }-
	\frac{1}{ (\omega_{b}-\omega_{a}-\Omega)_p }
	\right],
\end{split}
\end{equation}
where the subscript $p$ denotes principal value, $n_a$ and $n_b$ are the Bose-Einstein populations at harmonic frequencies $\omega_a$ and $\omega_b$. In contrast to the classical Lorentz-oscillator description, where damping enters as a single, $q$-independent phenomenological parameter, here the linewidth and line shape emerge from a fully $q$- and $\Omega$-dependent self-energy $\Sigma$, and need not be Lorentzian. An application of this spectral function to the calculation of black-body radiation in PbTe, where we find strong deviations from simple Lorentzian broadening and from Planck's law, underscoring the importance of a model-free treatment of broadening, is given in the End Matter.

As an important application we examine phonon polaritons in $\beta$-Ga$_2$O$_3$, which has been extensively studied experimentally \cite{Passler.2022,Matson.2023}. Specifically, the notion of hyperbolic polaritons has gained  interest for a variety of applications.\cite{Dai.2014,Galiffi.2024} A hyperbolic polariton denotes the appearance of a pattern that appears hyperbolic when a constant energy slice of the polariton dispersion curve is taken, in contrast with spherical or elliptical patterns expected for free photons.
To address this theoretically, in the left panel of Fig.\ \ref{fig:hyperbolic} we show the electromagnetic part of the phonon polaritons, defined as
\begin{equation}
    \label{eq:spectralEM}
    I(\vec{q},\Omega) = \sum_{j\mu} J_{\vec{q}jj}(\Omega) \nu_{\vec{q}j}^{\mu\,\dagger} \nu_{\vec{q}j}^{\mu},
\end{equation}
which is an integrated spectral function weighted by the norm of the electric field component, $\sum_\mu \nu_{\vec{q}j}^{\mu\,\dagger} \nu_{\vec{q}j}^{\mu}$. This spectral function reduces to the harmonic polaritonic dispersion curves in the absence of quasi-particle interactions. 
Complex patterns, including hyperbolic ones, emerge naturally if we examine constant energy cuts of this polariton spectral function, panels (i)-(viii). As can be observed, the patterns at constant energy cuts are an extremely rapid and complex function of energy, and all of them are captured here from first principles. Specifically, slice (i) compares well with the hyperbolic shear polaritons in Ref. \cite{Passler.2022},with our energy slice at an only slightly lower energy than experiment, 700 cm$^{-1}$ vs 718 cm$^{-1}$, an issue likely arising from the underlying DFT approximations.

Based on the above, we can propose a simple criterion for the emergence of hyperbolic polaritons --- the presence of a discontinuity at the zone center of the phonon branches, which must be ``bridged over'' by the polaritonic dispersion. This means, in practice, that any non-cubic polar material may exhibit this phenomenon. However, the energy range for its observation may vary widely. In GaN, as illustrated in Fig. \ref{fig:gan_polariton}non-trivial constant energy dispersions are expected only in the narrow energy window where there is a discontinuity, whereas in Ga$_2$O$_3$ there is an abundance of discontinuities, making the hyperbolic polaritons much easier to detect.

In conclusion, we presented non-empirical theory for phonon-polaritons in bulk materials. The key difference from established theory is that the electromagnetic field is explicitly included in the Hamiltonian as a dynamical variable. This has allowed us to compute harmonic and anharmonic  phonon-polariton dispersion relations, spectral functions, dielectric response, and black-body radiation in both simple and complex materials. Our treatment offers a comprehensive and predictive description of phonon-polaritons and sets the stage for exploring a variety of light-matter interaction phenomena. These features should become particularly relevant for polariton-mediated energy transport, nanophotonic control of heat in anisotropic crystals, and prospective extensions to polariton-assisted chemistry.

\section{Acknowledgements}

Supercomputer resources were provided by the National Academic Infrastructure for Supercomputing in Sweden (NAISS). L.K. acknowledges support from the Aryeh and Mintzi Katzman Professorial Chair and the Helen and Martin Kimmel Award for Innovative Investigation. 



\clearpage

\section{End matter}

In the End Matter, we revisit the polariton spectral function and compare its predictions to those obtained from ideal black-body theory, i.e, with the case of non-interacting photons and phonons.

\begin{figure}
    \includegraphics[width=\linewidth]{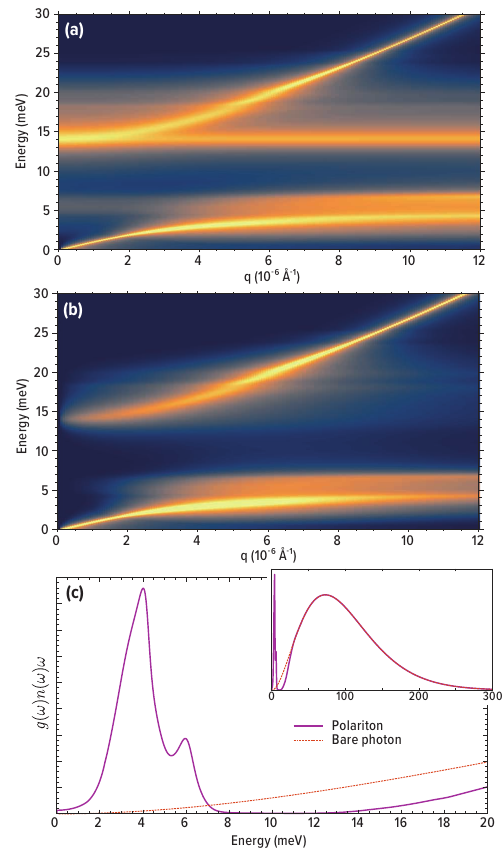}
    \caption{\label{fig:pbtespectral}
    (a) Polariton spectral functions for PbTe at 300K. Panel (b) is only the electromagnetic part of the polaritons, i.e. the photon spectral function. Panel (c) is the anharmonic photon density of states, compared with the one obtained using the ideal Planck's law.
    }
\end{figure}

The polariton spectral function contains both the lattice and electromagnetic components. As a benchmark, we consider PbTe in the rock-salt structure, known to have anomalous phonons at the zone 
center \cite{Li.2014}. 
In Fig.\ \ref{fig:pbtespectral}(a) we present the polariton spectral function, and in (b) we show only the photon-like parts of the spectrum (see Eq.\ (\ref{eq:spectralEM}) in the main text). The zone center phonons in PbTe deviate strongly from a simple Lorentzian broadening, a property ``inherited'' by the phonon polaritons in the form of a double-peak structure of the lower phonon polariton, seen in panel (b). Clearly, the effect of the strong phonon anharmonicity is not restricted to the phonon system, but also broadens the photon-like parts into complex line shapes. We note in passing that the contribution from the three-phonon anharmonicity is, by several orders of magnitude, the dominant term in determining the broadening. The terms in Eq.\ (\ref{eq:anharmonicterms}) that involve electric fields have a negligible contribution to the self-energy. This can be inferred from the fact that the integral in Eq.\ (\ref{eq:selfenergy}) picks up anharmonic contributions from the entire Brillouin zone, but the electromagnetic contributions are confined to a small fraction of the domain.

By integrating the electromagnetic part of the spectral function, we can define the photon density of states as
\begin{equation}
    g(\Omega) = 
    \sum_{\vec{q}j} -\frac{1}{\pi} 
    \Im \left\{ G_{\vec{q}jj}(\Omega) \right\}
    \sum_{\mu} 
    \nu_{\vec{q}j}^{\mu \dagger} \nu_{\vec{q}j}^\mu,
\end{equation}
where the sum over $\mu$ encompasses the EM part of the polariton eigenvectors. We can then obtain the black-body spectrum, shown in Fig. \ref{fig:pbtespectral}c. At frequencies far higher than the phonon frequencies, the photon density of states scales as $\omega^2$ and we recover Planck's law of thermal radiation \cite{Planck.1901}, as shown in the inset of Fig. \ref{fig:pbtespectral}(c). In the polariton regime, however, there is a strong deviation from Planck's law. The absorption energy range is consistent with the fact that lead chalcogenides are used as THz detectors \cite{Han.2016,Komissarova.2007}.

The capability to determine the deviation from an ideal black-body spectrum from first principles opens up the possibility to study emissivity and thermal radiation.

\end{document}